\documentclass[12pt]{article} 

\usepackage{graphicx}
\usepackage{amsmath}
\usepackage{amssymb}
\usepackage{hyperref}
\usepackage[height=8.8in,width=6.45in]{geometry}
\usepackage{tikz}

\numberwithin{equation}{section}

\newcommand{\UU}{\mathrm{U}}
\newcommand{\OO}{\mathrm{O}}
\newcommand{\SU}{\mathrm{SU}}

\newcommand{\bC}{\mathbb{C}}
\newcommand{\CC}{{\mathbb C}}
\newcommand{\CP}{{\CC\PP}}

\newcommand{\PP}{{\mathbb P}}

\newcommand{\bR}{\mathbb{R}}
\newcommand{\R}{\bR}

\newcommand{\WW}{{\mathbb W}}
\newcommand{\bZ}{\mathbb{Z}}
\newcommand{\ZZ}{{\mathbb Z}}

\newcommand{\cE}{\mathcal{E}}
\newcommand{\cL}{\mathcal{L}}

\newcommand{\cN}{\mathcal{N}}
\newcommand{\cO}{\mathcal{O}}

\newcommand{\cQ}{{\cal Q}}

\newcommand{\fg}{\mathfrak{g}}

\newcommand{\fM}{\mathfrak{M}}
\newcommand{\fR}{\mathfrak{R}}

\newcommand{\bphi}{\overline{\phi}}

\newcommand{\dd}{{\rm d}}
\newcommand{\e}{\mathrm{e}}
\newcommand{\mc}{{\rm g}}

\newcommand{\trans}{{\sf T}}

\DeclareMathOperator{\tr}{tr}
\DeclareMathOperator{\Tr}{Tr}
\DeclareMathOperator{\im}{\mathbb{I}m}
\DeclareMathOperator{\re}{\mathbb{R}e}
\DeclareMathOperator{\rank}{rank}

\newcommand{\eg}{\textit{e.g.}}
\newcommand{\ie}{\textit{i.e.}}

\newcommand{\beq}{\begin{equation}}
\newcommand{\eeq}{\end{equation}}
\newcommand{\be}{\begin{equation}}
\newcommand{\ee}{\end{equation}}
\newcommand{\bea}{\begin{equation}\begin{aligned}}
\newcommand{\eea}{\end{aligned}\end{equation}}

\newcommand{\nn}{\nonumber}

\newcommand{\intd}[2]{{\raisebox{#1}{\scriptsize \ensuremath{#2}}}}

\begin{document}

\begin{titlepage}

\begin{flushright}
IPMU-13-0082\\
UT-13-17\\
\end{flushright}

\vskip 3cm

\begin{center}
{\Large \bf
Elliptic genera of two-dimensional \\[.7cm]
$\cN=2$  gauge theories  with rank-one gauge groups
}

\vskip 2.0cm

Francesco Benini$^\sharp$,
Richard Eager$^\natural$,
Kentaro Hori$^\natural$,
and Yuji Tachikawa$^\flat$
\bigskip
\bigskip

\begin{tabular}{ll}
$^\sharp$&Simons Center for Geometry and Physics, Stony Brook University,\\
&Stony Brook, NY 11794, USA\\
$^\natural$  & Institute for the Physics and Mathematics of the Universe, \\
& University of Tokyo,  Kashiwa, Chiba 277-8583, Japan\\
$^\flat$  & Department of Physics, Faculty of Science, \\
& University of Tokyo,  Bunkyo-ku, Tokyo 133-0022, Japan
\end{tabular}

\vskip 1cm

\textbf{Abstract}
\end{center}

\medskip
\noindent
We compute the elliptic genera of two-dimensional $\cN=(2,2)$ and $\cN=(0,2)$ gauge theories via supersymmetric localization, for rank-one gauge groups.
The elliptic genus is expressed as a sum over residues of a meromorphic function whose argument is the holonomy of the gauge field along both the spatial and the temporal directions of the torus.
We illustrate our formulas by a few examples including the quintic Calabi-Yau, $\cN=(2,2)$ $\SU(2)$
and $\OO(2)$ gauge theories coupled to $N$ fundamental chiral multiplets, and a geometric $\cN=(0,2)$ model.

\bigskip
\vfill
\end{titlepage}

\setcounter{tocdepth}{2}
\tableofcontents

\section{Introduction and summary}

In the last few years we learned how to compute the partition function of various supersymmetric theories on products of spheres via localization.%
\footnote{There are many papers and we cannot  cite all of them. Interested readers should consult the inspire-hep database \cite{inspire}.}
The simplest of all cases is a two-dimensional supersymmetric theory on the torus $S^1\times S^1.$  The aim of this paper is to find a formula for the partition function of the system, which is called the \emph{elliptic genus}, when the theory has $\cN=(2,2)$ or $\cN=(0,2)$ supersymmetry.
Recent developments in localization computations were partially based on an improved understanding of rigid supersymmetry on curved manifolds, with spheres as prime examples.  The torus is flat, and therefore we do not need to use any of the recent technical advances.  In fact, this paper only requires techniques that were known twenty years ago.

Historically, the elliptic genus of free orbifolds was studied in \cite{Schellekens:1986yj,Schellekens:1986yi,Pilch:1986en},
and that of the non-linear sigma models was studied in \cite{Witten:1986bf,Witten:1987cg}.
The elliptic genus of Gepner models  was  computed using the known characters of $\cN=2$ superconfomal algebras \cite{Eguchi:1988vra,Ooguri:1989fd}. Then it was realized that the elliptic genus of Landau-Ginzburg models can be computed by localization \cite{Witten:1993jg}, which led to a formula for the elliptic genus of Gepner models using the orbifold Landau-Ginzburg description \cite{Berglund:1993fj,Kawai:1993jk,Kawai:1994fm,Berglund:1994zg}. The elliptic genus of a subspace of a K\"ahler quotient was studied in \cite{Eguchi:1999jm,Ando:2009av}. There,  the free field expressions for the fields on a K\"ahler quotient were used, but the elliptic genus of the ambient K\"ahler quotient  was computed by a known geometric formula.  The goal of the present paper is to provide a path-integral derivation of the elliptic genus.  The essential point of our paper is to carefully study the contribution from the zero-modes of the  gauge multiplets.  When the theory under consideration has a smooth geometric phase, our formula reproduces known mathematical results of the elliptic genus of complete intersections in toric varieties \cite{MaZhou, GuoZhou}.

Let us first consider a $\UU(1)$ gauge theory as an example.  The most important zero-modes are the holonomy of  the gauge group on the spacetime torus $T^2$. We denote the holonomy by
\be
u = \oint A_t \, \dd t - \tau \oint A_s \, \dd s
\ee
where $A$ is the gauge field, $\tau$ is the complex structure of the torus, and $t,s$ are temporal and spatial directions. As we have the identification $u\sim u+1 \sim u+\tau$,
the variable $u$ also takes values in $T^2$.
Very naively, localization would give rise to an integral
\be
\label{foo}
Z_{T^2}=\int \dd u \, \dd\bar u \; \tilde Z_\text{1-loop}(u,\bar u)
\ee
where $\tilde Z_\text{1-loop}$ is the one-loop determinant of the fluctuations around a given background gauge field.
It turns out that the naive one-loop determinant diverges at a finite number of points $u\in T^2$ and the integral \eqref{foo} needs to be carefully defined. In the end we will instead find a formula of the form
\be
\label{bar}
Z_{T^2} = - \sum\nolimits_{i} \oint_{u\,=\,u_i} \dd u  \; Z_\text{1-loop}(u)
\ee
where $Z_\text{1-loop}(u)$ is a meromorphic function on $T^2$ and the sum is over only a specific set of poles.%
\footnote{Such a formula was already given in \cite{Grassi:2007va}, where however the authors did not derive  how to choose the poles to sum over.}
We will derive below the general formula for $Z_\text{1-loop}(u)$ and the method to choose the correct set of poles, from a careful analysis of the path integral.

The formula for the partition function of a $d$-dimensional supersymmetric theory on $S^1\times S^{d-1}$ is superficially similar to the formula \eqref{bar} above, except that the integration contour $\int_{u=0}^{u=1}$ is used instead. In that case, $Z_\text{1-loop}(u)$ counts gauge-dependent words constructed out of basic letters, and the integral serves to project them down to gauge invariant operators.

The rest of the paper is organized as follows. In Sec.~\ref{sec: one-loop} we recall the basic definition of the elliptic genera, the structure of the supersymmetry multiplets  and the one-loop determinants. In Sec.~\ref{sec: rank-one} we obtain a general formula for the elliptic genera of rank-one gauge theories.
In Sec.~\ref{sec: examples} we discuss a few examples with $\cN=(2,2)$ supersymmetry, including the quintic Calabi-Yau as well as $\SU(2)$ and $\OO(2)$
gauge theories with massless flavors, and one example with $\cN=(0,2)$ supersymmetry.
We have three appendices: in App.~\ref{app: math} we give the mathematical formula of the elliptic genera of subvarieties of K\"ahler quotients;  App.~\ref{app: theta} collects our conventions on $\eta$ and $\theta$ functions; App.~\ref{app: actions} collects supersymmetry transformation laws and actions for $\cN=(2,2)$ and $\cN=(0,2)$ theories.

We will generalize this discussion to higher-rank gauge groups in a forthcoming paper \cite{toappear}.

\paragraph{Note added:} The authors thank Abhijit Gadde and Sergey Gukov for notifying them of an  imminent submission of a related paper, and for kindly giving them an additional ten days to prepare this paper.

\section{Elliptic genera, multiplets, one-loop determinants}
\label{sec: one-loop}

We begin by presenting the definition of the elliptic genus of two-dimensional theories with $\cN=(2,2)$ and $\cN=(0,2)$ supersymmetry. We will also compute the one-loop determinants of quadratic fluctuations around a background with flat connections, which is equivalent to a trivial background but with non-trivial boundary conditions.

\subsection{Theories with $\cN=(2,2)$ supersymmetry}

Let us consider a two-dimensional $\cN=(2,2)$ theory with flavor symmetry group $K$ (with Cartan generators $K_a$) and a left-moving $\UU(1)$ R-symmetry $J$ (which is discrete if the theory is not conformal).
Its elliptic genus is defined as
\be
\label{def genus 2,2}
Z_{T^2}(\tau, z, u) =
\Tr_\text{RR} \, (-1)^F q^{H_L} \bar q^{H_R} y^J \prod\nolimits_a x_a^{K_a} \;,
\ee
where the trace is taken in the RR sector which means that fermions have periodic boundary conditions. Here $F$ is the fermion number,
\be
\label{def q}
q = e^{2\pi i \tau}
\ee
specifies the complex structure of a torus and we write $\tau = \tau_1 + i \tau_2$. The left- and right-moving Hamiltonians
$H_L$ and $H_R$ are $2H_L = H + i P$, $2H_R = H - i P$ in Euclidean signature. Since $q^{H_L} \bar q^{H_R} = \exp( - 2\pi \tau_2 H - 2\pi \tau_1 P)$, the trace can be represented by a path integral on a torus of complex structure $\tau$. For a superconformal theory, the operators $H_L, H_R, J$ equal the zero-mode generators $L_0, \bar L_0, J_0$ of the superconformal algebra.%
\footnote{When not uniquely fixed, \eg{} by the superpotential, the superconformal R-symmetries can be determined through the $c$-extremization principle of \cite{Benini:2012cz, Benini:2013cda}.}
We also define
\be
\label{def y x}
y = e^{2\pi i z} \qquad\quad\text{and}\qquad\quad x_a = e^{2\pi i u_a}
\ee
which specify the background R-symmetry and flavor-symmetry gauge fields $A^\text{R}$, $A^\text{flavor}$ via
\be
z = \oint_t A^\text{R} - \tau \oint_s A^\text{R} \;,\qquad
u_a =\oint_t A^\text{$a$-th flavor} - \tau \oint_s A^\text{$a$-th flavor} \;,
\ee
where $t,s$ are the temporal and spatial cycles.%
\footnote{In particular introducing a complex coordinate $w \sim w+1 \sim w + \tau$ on the torus, the relation between $z$ and a constant connection $A^\text{R}_\mu$ is $z = (-2i \tau_2)\, A^\text{R}_{\bar w}$ (and similarly for the flavor holonomies).}
Equivalently we can set the background gauge fields to zero and instead specify twisted boundary conditions.
When the R-symmetry is discrete, $z$ is only allowed to take certain discrete values.
When $z=u=0$ the elliptic genus reduces to the Witten index, and when the 2d theory has a smooth geometric description it gives the Euler number of the target manifold.
The limit $q\to 0$ of the elliptic genus is called the $\chi_y$ genus.

We concentrate on gauge theories with vector and chiral multiplets, possibly with superpotential and twisted superpotential interactions. Our conventions are given in Appendix \ref{app: actions}.
A chiral multiplet has components $\Phi = (\phi, \bar\phi, \psi, \bar\psi, F, \bar F)$ and a vector multiplet has $V = (A_\mu,\sigma, \bar\sigma, \lambda,\bar\lambda, D)$.

The computation of one-loop determinants on a background where only flat connections and D-terms are turned on, involves the evaluation of infinite products of the form
\be
\label{fermionproduct}
\prod_{m,n} (m+n\tau+u)
\ee
for left-moving fermions, its complex conjugate for right-moving fermions, and
\be
\label{bosonproduct}
\prod_{m,n} \frac{1}{|m+n\tau+u|^2 + i D}
\ee
for bosons. Here $u$ and $D$ denote  the holonomies and the D-component%
\footnote{As explained in Appendix \ref{app: actions}, $D$ is the Eulidean D-term related to the Lorentzian one by $D_L = i D$. Moreover in (\ref{bosonproduct}) we have rescaled $D$ by $\pi/\tau_2$.}
of background vector multiplets. In isolation, the products \eqref{fermionproduct} and \eqref{bosonproduct} require renormalization by infinite constants, but with $\cN=(2,2)$ supersymmetry the constants simply cancel out.

\paragraph{Chiral multiplet.}
The contribution from a chiral multiplet $\Phi$ of vector-like R-charge%
\footnote{The definition (\ref{def genus 2,2}) contains the left-moving R-charge $J$. A chiral multiplet of vector-like R-charge $R$ (and assigning vanishing axial R-charge) has $J = \frac R2$.}
$R$ and flavor charge $Q$ is
\be
\label{eq:chiralproduct}
Z_{\Phi,Q}(\tau,z,u,D) = \prod_{m,n} \frac{\big( m+n\tau+(1- \tfrac R2 ) z - Q u \big) \big( m + n \bar\tau + \frac R2 \bar z + Q \bar u \big)}{ \big| m + n\tau + \frac R2 z + Qu \big|^2 + i QD}
\ee
which, when $D=0$, simplifies to
\be
\label{chiral}
Z_{\Phi,Q}(\tau,z,u)=\frac{\theta_1(q,y^{R/2-1} x^{Q}) }{\theta_1(q,y^{R/2} x^Q)} \;.
\ee
Here and in the following we will use interchangeably $\tau,z,u$ and $q,y,x$ using the relations (\ref{def q}) and (\ref{def y x}).
Notice that $Z_{\Phi,Q}(\tau,z,u,D)$ is meromorphic in $D$ but not in $\tau,z, u$, while $Z_{\Phi,Q}(\tau,z,u)$ is a meromorphic function of its arguments.
The function $\theta_1(q,y)$, that we also denote as $\theta_1(\tau|z)$, is a Jacobi theta-function and it is defined in Appendix \ref{app: theta}. The infinite product has been regularized in such a way to match the Hamiltonian computation. The formula \eqref{chiral} was derived twenty years ago in \cite{Witten:1993jg}. There, the Landau-Ginzburg model with superpotential $W=\Phi^{k+2}$ was considered, and therefore $\Phi$ had R-charge $2/(k+2)$. With this value and $x=1$, the formula \eqref{chiral} reproduces the elliptic genus of the $k$-th minimal model.

The contribution of a chiral multiplet in a general representation $\fR$ of a gauge group $G$ is, when $D=0$:
\be
\label{chiral rep}
Z_{\Phi,\fR}(\tau, z, u)
= \prod_{\rho \,\in\,\fR}
\frac{\theta_1(q, y^{R/2-1} x^{\rho})}{\theta_1(q, y^{R/2} x^\rho)}
\ee
where the product is over the weights $\rho$ of the representation,
and $x^\rho \equiv e^{2\pi i\rho(u)}$.

\paragraph{Vector multiplet.}
The contribution from a vector multiplet of a group $G$ consists of two parts. The Cartan part, with the zero-modes removed, contributes a factor
\be
\label{vector Cartan}
Z_\text{vect}(\tau,z)^r  \qquad\text{where}\qquad  Z_\text{vect}(\tau,z)= \frac{i \eta(q)^3}{\theta_1(q,y^{-1})}
\ee
where  $r$ is the rank of $G$. In this paper we will consider $r=1$. Note that the numerator $i\eta(q)^3$ equals
$ \lim_{y\to 1} (1-y)^{-1}\theta_1(q,y)$.
The off-diagonal components give
\be
Z_\text{off}(\tau,z,u,D)
= \prod_{\alpha:\text{ roots}} \,
\prod_{m,n} \frac{\big( m+n\tau  - \alpha(u) \big) \big(m + n\bar\tau
- \bar z + \alpha(\bar u) \big)}{\big| m+n\tau - z +
\alpha(u) \big|^2 + i \alpha(D)} \;,
\ee
where $u$ and $D$ denote the background gauge holonomy and
the D-component vacuum expectation value in the Cartan subgroup.
The product is over the roots of the gauge group.
The dependence on $z$ is fixed by the fact that the boson and
the right-moving fermions have left-moving R-charge $-1$,
while the left-moving fermion is uncharged.
When $D=0$ the formula simplifies to
\be
\label{vector off}
Z_\text{off}(\tau,z,u)=\prod_{\alpha: \text{ roots}}\frac{\theta_1(q,x^{\alpha}) }{\theta_1(q,y^{-1} x^\alpha)}.
\ee
Notice that this is the same contribution as of a twisted chiral multiplet of axial R-charge $2$ and vanishing vector-like R-charge.

\paragraph{Twisted chiral multiplet.}
The one-loop determinant of a twisted chiral multiplet of axial R-charge $R_A$ can be similarly computed:
\be
\label{twisted chiral}
Z_\Sigma = \frac{\theta_1(q, y^{-R_A/2 + 1})}{\theta_1(q, y^{-R_A/2})} \;.
\ee
As a check, the twisted chiral multiplet $\Sigma = (\sigma, \lambda, F_{12} + iD)$ containing the field strength has $R_A = 2$ and it gives us the vector multiplet determinant.

\subsection{Theories with $\cN=(0,2)$ supersymmetry}

For two-dimensional theories with $\cN=(0,2)$ supersymmetry the situation is similar. Let $K$ be the flavor symmetry group. The elliptic genus is defined as
\be
Z_{T^2}(\tau, u) = \Tr_\text{RR} (-1)^F q^{H_L} \bar q ^{H_R}
\prod\nolimits_a x_a^{K_a}
\ee
where, again, $q = e^{2\pi i \tau}$ and $x_a = e^{2\pi i u_a}$. Lagrangian theories can be written in terms of three types of multiplets. The chiral multiplet $\Phi = (\phi, \psi^-)$ consists of a complex scalar $\phi$ and a right-moving Weyl fermion $\psi^-$. The Fermi multiplet $\Lambda^+ = (\psi^+, G)$ consists of a left-moving Weyl fermion $\psi^+$ and an auxiliary complex scalar $G$; the definition of this multiplet involves a chiral multiplet $(E, \psi^-_E)$ where $E(\phi)$ is a holomorphic function of the fundamental chiral multiplets in the theory---see Appendix \ref{app: actions}. The vector multiplet $V = (A_\mu, \lambda^+, D)$ consists of a vector $A_\mu$, a left-moving Weyl fermion $\lambda^+$ and an auxiliary real scalar $D$.

As before, the contributions from the three types of multiplets can be written in terms of (\ref{fermionproduct}) and (\ref{bosonproduct}). In this case the expressions in isolation are not well-defined and require $\tau$-dependent renormalization. Even the product of all one-loop determinants in the theory requires renormalization whenever there is a gravitational anomaly. One could cure the problem simply by multiplying and dividing by the one-loop determinants of free chiral or Fermi multiplets in a suitable number to cancel the gravitational anomaly. Alternatively we can compute the one-loop determinants of free fields in the Hamiltonian formalism---precisely as in \cite{Witten:1993jg}.

\paragraph{Chiral and Fermi multiplets.}
The contributions from a chiral multiplet of charge $Q$, on a background with $D\neq0$ and $D=0$ respectively, are
\be
Z_{\Phi,Q}(\tau,u,D) = \prod_{m,n} \frac{m + n \bar\tau + Q \bar u}{|m + n\tau + Qu|^2 + iQD} \;,\qquad\qquad Z_{\Phi,Q}(\tau,u) = i\, \frac{ 
\eta(q)}{\theta_1(q, x^Q)} \;.
\ee
The contribution from a Fermi multiplet of charge $Q$ is
\be
Z_{\Lambda,Q}(\tau,u) = i\, \frac{\theta_1(q, x^Q)}{
\eta(q)} \;,
\ee
with no dependence on $D$. For general representation $\fR$, we simply multiply over weights $\rho \in \fR$ as in (\ref{chiral rep}). Note that the $q$-expansions can start with a nontrivial power $q^E$, where $E$ is the Casimir energy of the multiplet.

It is easy to check that the product of the determinants of a chiral multiplet with left-moving R-charge and flavor charge $\big( \frac R2,Q \big)$, and of a Fermi multiplet with charges $\big( \frac R2-1,Q \big)$, reproduces the determinant of an $\cN=(2,2)$ chiral multiplet in (\ref{chiral}) (up to a sign which depends on a different definition of the fermion number). Moreover the product of the determinants of a chiral and a Fermi multiplet of opposite charge is 1, since they can be given a supersymmetric mass and be integrated out.

\paragraph{Vector multiplet.} The contributions from a vector multiplet corresponding to a Cartan generator (with the zero-modes removed) and a root $\alpha$ of the gauge group are
\be
Z_\text{vect}(\tau) = 
\eta(q)^2 
\;,\qquad\qquad\qquad Z_\text{off}(\tau, u) = i\, \frac{\theta_1(q, x^\alpha)}{
\eta(q)} \;.
\ee
Notice that the determinant of an off-diagonal vector multiplet is exactly equal to that of a Fermi multiplet, since in two dimensions the gauge field is non-dynamical and thus the two contain the same degrees of freedom. Moreover the product of the determinants of a vector (or Fermi) multiplet of left-moving  R-charge $0$ and of a chiral multiplet of left-moving  R-charge $-1$, reproduces the determinant of an $\cN=(2,2)$ vector multiplet in (\ref{vector Cartan}) and (\ref{vector off}).

\

The analysis in the next section is focused on the $\cN=(2,2)$ case, but it is equally applicable to $\cN=(0,2)$ theories, and we will give a concrete example in Section \ref{sec:DK}.

\section{The formula and its path-integral derivation}
\label{sec: rank-one}

We will now present our formula for the elliptic genus, in case the gauge group has rank one, and its
derivation. In a forthcoming paper we will generalize it to higher-rank gauge groups.

\subsection{The setup and the result}

Let us first consider an $\cN=(2,2)$ $\UU(1)$ gauge theory with chiral multiplets $\Phi_i$, of $\UU(1)$ charges $Q_i$ and R-charges $R_i$, where $i$ runs over all chiral multiplets in the theory. Without superpotential the theory has some flavor symmetry group, while a non-trivial superpotential $W$ in general reduces such flavor symmetry. We will denote by $G_F$ the flavor symmetry group, and by $P_i$ the flavor charges under the maximal torus of $G_F$.
We assume that we can arrange $R_i$ to be all positive,
\be
\label{assumption R-charges}
R_i>0 \;,
\ee
if we use the freedom to shift $R_i$ by gauge or flavor charges.
We show in section \ref{sec: u1derivation} that the elliptic genus is given by
\be
\label{u1formula}
Z_{T^2}(\tau,z,\xi)
= - \sum_{u_j \,\in\, \fM^+_\text{sing}}
\oint_\intd{-2pt}{u=u_j} \hspace{-1.2em} \dd u \;
\frac{i\eta(q)^3}{\theta_1(q,y^{-1})}
\prod_{\Phi_i} \frac{\theta_1 \big( q \,,\, y^{R_i/2-1} \, x^{Q_i} \,
e^{2\pi i P_i(\xi)} \big) }{\theta_1 \big( q \,,\,
y^{R_i/2} \, x^{Q_i} \, e^{2\pi i P_i(\xi)} )} \;.
\ee
$\xi$ specifies the holonomy of the flavor symmetry on the torus.
Here, $u$ parameterizes the gauge holonomy and takes values in the torus
$\fM = \bC/(\bZ+\tau \bZ)$.
When the gauge theory is not conformal $z$ is chosen so that the integrand is periodic under the identification $u\sim u+1\sim u+\tau$. Note that
$Z_{T^2}(\tau, z, \xi)$ might develop poles as $\xi$ is
varied on the maximal torus, so we will assume that $\xi$ is generic.

The poles to sum over, $\fM^+_\text{sing}$, are chosen as follows. There are poles in the integrand at
\be
\label{sing}
Q_i u + \frac{R_i}2 \,z + P_i(\xi) = 0 \pmod{\bZ + \tau\bZ} \qquad\quad \text{for some $i$}
\ee
where the chiral multiplet $\Phi_i$ is massless. We shall denote the set of such points by $\fM_\text{sing}$. It may well happen that at a point $u_* \in \fM_\text{sing}$ more than one multiplet is massless.
Under the assumption (\ref{assumption R-charges}), for generic values of $z$ the massless multiplets at $u_*$ have charges $Q_i$ all with the same sign.%
\footnote{\label{foot: R-charges}
If (\ref{assumption R-charges}) is not satisfied, as it happens in the non-Abelian case, it might happen that the splitting in (\ref{splitting of singularities}) is not well-defined. We first compute the elliptic genus in the absence of superpotential $W$: in this case it is always possible to separate  the singularities so that (\ref{splitting of singularities}) is well-defined. In other words it is always possible, by mixing the R-symmetry with gauge and flavor symmetries, to achieve (\ref{assumption R-charges}). Then we analytically continue the result to the R-charges allowed by the superpotential.}
Then we split the points in $\fM_\text{sing}$ in two groups according to $Q_i>0$ or $Q_i<0$:
\be
\label{splitting of singularities}
\fM_\text{sing} = \fM_\text{sing}^+ \sqcup \fM_\text{sing}^- \;.
\ee
This determines the set $\fM_\text{sing}^+$ in (\ref{u1formula}).

In (\ref{u1formula}) we could have equivalently taken the positive sum over $u_j \in \fM_\text{sing}^-$. Our formula is then
\be
\label{u1formula neg}
Z_{T^2}(\tau,z,\xi)
= \sum_{u_j \,\in\, \fM^-_\text{sing}} \oint_\intd{-2pt}{u=u_j}
\hspace{-1.2em} \dd u \; \frac{i\eta(q)^3}{\theta_1(q,y^{-1})}
\prod_{\Phi_i}
\frac{\theta_1 \big( q \,,\, y^{R_i/2-1} \, x^{Q_i} \,
e^{2\pi i P_i(\xi)} \big) }{\theta_1 \big( q \,,\, y^{R_i/2}
\, x^{Q_i} \, e^{2\pi i P_i(\xi)} )} \;.
\ee
This is equal to \eqref{u1formula}, as the sum of the residues of a meromorphic function on a torus is zero.

In Section \ref{sec: examples} we present some illustrative examples: the quintic Calabi-Yau, a singular hypersurface Calabi-Yau in a toric variety, and $\CC\PP^N$. Readers interested in these examples can consult that section before proceeding.

\

Similarly, the elliptic genus of a gauge theory with gauge group $G$ of rank one can be derived with a minimal modification. There is an additional factor (\ref{vector off}) from the off-diagonal components of the gauge multiplet in the integrand of (\ref{u1formula}), and an overall factor $1/|W|$, where $|W|$ is the order of the Weyl group, to account for its identifications. Each chiral multiplet $\Phi_i$ has vector-like R-charge $R_i$ and it is in the representation $\fR_i$ of $G$. When $G$ is connected, we find:
\begin{multline}
\label{Gformula}
Z_{T^2}(\tau,z,\xi)
= - \frac1{|W|} \sum_{u_j \,\in\, \fM^+_\text{sing}} \oint_\intd{-2pt}{u=u_j} \hspace{-1em} \dd u\, \frac{i\eta(q)^3}{\theta_1(q, y^{-1})} \prod_{\alpha \,\in\, G} \frac{\theta_1(q, x^{\alpha})}{\theta_1(q, y^{-1} x^\alpha)} \times\phantom{,} \\
\phantom{,}\times \prod_{\Phi_i} \prod_{\rho\,\in\,\fR_i} \frac{\theta_1 \big(q \,,\, y^{R_i/2-1} \, x^{\rho} \, e^{2\pi i P_i(\xi)} \big)}{\theta_1 \big( q \,,\, y^{R_i/2} \, x^\rho \, e^{2\pi i P_i(\xi)} \big)} \;.
\end{multline}
In Section \ref{sec: SU2} we present the example of $\SU(2)$ with $N$ fundamentals.
The derivation described below can also be applied to the case where $G$
is not connected. In Section \ref{sec:O2} we illustrate this in a particular case, where $G=\OO(2)$.

\

Finally, the elliptic genus of a  theory with $\cN=(0,2)$ supersymmetry, gauge group $G$, chiral multiplets $\Phi_i$ and Fermi multiplets $\Lambda_j$, is given by
\begin{multline}
\label{02formula}
Z_{T^2}(\tau,\xi) = - \frac1{|W|} \sum_{u_j \,\in\, \fM_\text{sing}^+} \oint_\intd{-2pt}{u = u_j} \hspace{-1em} \dd u\; 
\eta(q)^2
\, \prod_{\alpha\,\in\,G} \frac{i\,\theta_1(q,x^\alpha)}{
\eta(q)} \times\phantom{,} \\
\phantom{,}\times \Bigg( \prod_{\Phi_i} \prod_{\rho\,\in\,\fR_i} \frac{i \, 
\eta(q)}{\theta_1(q, x^\rho)} \Bigg) \Bigg( \prod_{\Lambda_j} \prod_{\rho \,\in\, \fR_j} \frac{i\, \theta_1(q, x^\rho)}{
\eta(q)} \Bigg) \;.
\end{multline}
The authors of \cite{Kawai:1994np} observed that the elliptic genus of a rank-one Abelian $\cN=(0,2)$ GLSM precisely takes this form.  They asked if there is a path-integral derivation.  The answer is yes as we will now explain.

\subsection{Derivation}
\label{sec: u1derivation}

In this section we provide a path-integral derivation of the formula \eqref{u1formula} for the elliptic genus.
The generalization to (\ref{Gformula}) and (\ref{02formula}) is straightforward.
We denote by $\e$  the gauge coupling, so that we have a factor $1/\e^2$ in front of the gauge kinetic terms.
Similarly, we put $1/\mc^2$ in front of the  matter kinetic terms.
We consider the localization as $\e\to 0$ and
$\mc\to 0$. This is possible because (as reviewed in Appendix \ref{app: actions}) all Lagrangian terms are actually $\cQ$-exact. This implies that we can take the kinetic terms much larger than the interaction terms, and moreover the result will not have dependence on the parameters in the interaction terms.
In what follows, some $\tau_2$-dependent rescalings of fields and
parameters are assumed.

For non-zero couplings $\e$ and $\mc$, the elliptic genus can be written as
\be
\label{int0}
Z_{T^2} = \int_\R \dd D \int_{\fM } \dd^2u \,
f_{\e,\mc}(u,\bar u,D) \, \exp\Big[ - \frac1{2\e^2}D^2 - i \zeta D \Big] \;.
\ee
where $f_{\e,\mc}(u,\bar u,D)$ is the result of the path integral except the vector multiplet zero-modes.
Here $\zeta$ is a Fayet-Iliopoulos term that we include for completeness, but the result does not depend on it.

\subsubsection{Identification of dangerous regions}
\label{sec: dangerous}

Let us identify the dangerous regions when we take $\e\to 0$ and $\mc\to 0$.
After integrating over $D$ in \eqref{int0}, we have
\be
\label{int}
Z_{T^2} = \int_\fM \dd^2u\, F_{\e,\mc}(u,\bar{u}) \;.
\ee
The function $F_{\e,\mc}(u,\bar u)$ depends on $\e$ and $\mc$, but the dependence disappears after taking
the integral in (\ref{int}).
Consider the limit of $F_{\e,\mc}(u, \bar{u})$  as $\e\to 0$
and/or $\mc\to 0$. The sources of danger are the scalar zero-modes,
which exist at $u \in \fM_\text{sing}$ defined in \eqref{sing}. As we commented before (\ref{splitting of singularities}) and in footnote \ref{foot: R-charges}, at generic values of $z$ we can assume that
at any value of $u$ only  fields of the same sign of  $\UU(1)$ charges can simultaneously develop zero-modes.
Therefore the $\mc\to 0$ limit exists for any $u$
as long as $\e$ is non-zero, because the quartic potential induced by the D-term gives a cutoff.

Let us estimate the behavior of $F_{\e,0}(u, \bar{u})$ as $\e \to 0$
near a singular point $u_*\in \fM_\text{sing}$.
Suppose that there are $M_*$ fields
that have a zero-mode $\phi_i$ at $u_*$, and for simplicity let us assume that all have the same charge $Q$.
Then, by performing the $D$-integral in \eqref{int0}, we find that for $u\sim u_*$
\be
F_{\e,0}(u, \bar{u}) = C_{u,\e} \int_{\bC^{M_*}}\dd^{2M_*}\phi\,
\exp \bigg[ -\frac{1}{\mc^2}\big| Q(u-u_*) \big|^2|\phi_i|^2
- \frac{\e^2}2 \big( Q |\phi_i|^2 - \zeta\big)^2 \bigg] \;.
\ee
Here,  $C_{u,\e}$ comes from the non-zero modes and has a limit as $u\to u_*$ and $\e\to 0$.
The rest of the integral comes from the almost-zero modes.
The exponent in the integrand has two terms, the first coming from the kinetic term of the chiral multiplets
and the second coming from the quartic interaction which is the square of the D-term.

We see that $F_{\e,0}(u, \bar{u})$ has a finite limit as $\e\to 0$ provided $u \ne u_*$.
On the other hand for any $u\sim u_*$, including $u=u_*$, it is bounded as
\be
\label{bound}
\big| F_{\e,0}(u, \bar{u}) \big| \leq |C_{u_*,0}|  \Big( \frac C{|Q|\, \e} \Big)^{M_*} \;,
\ee
for some positive constant $C$ independent of $\e$.

Let $\Delta_{\varepsilon}$ be the $\varepsilon$ neighborhood of
$\fM_\text{sing}$ in $\fM$.
We may allow the radius to depend on the component, $\varepsilon=\varepsilon_*$ at $u_*$.
We separate the integral as
\be
Z_{T^2} = \int_{\fM\setminus \Delta_{\varepsilon}} \dd^2u\, F_{\e,0}(u, \bar{u})
+ \int_{\Delta_{\varepsilon}} \dd^2u\, F_{\e,0}(u, \bar{u}) \;.
\ee
We shall evaluate this in the limit $\e\to 0$, and we would like to do
so in such a way that the second term does not contribute.
In view of (\ref{bound}), we cannot take the limit $\e\to 0$ before
$\varepsilon\to 0$. We can take the limit $\varepsilon \to 0$
first for a non-zero $\e$ and then $\e\to 0$, or alternatively
we can take the scaling limit $\varepsilon\to 0$ and $\e\to 0$ in which the bound
\be
\varepsilon_* < \e^{M_*+1}
\ee
is satisfied. Denoting such a limit by
\raisebox{0pt}[0pt][0pt]{$\displaystyle \lim_{\e, \varepsilon \to 0}$}, we have
\be
Z_{T^2} =
\lim_{\e,\varepsilon\to0} \int_{\fM\setminus \Delta_\varepsilon} \dd^2u\, F_{\e,0}(u, \bar{u}) \;.
\ee

\subsubsection{Reintroduction of the auxiliary field $D$}

Now let us consider the path integral before integrating out the auxiliary field $D$:
\be
\label{int2}
Z_{T^2} = \lim_{\e,\varepsilon\to0} \int_\R \dd D
\int_{\fM\setminus \Delta_{\varepsilon}} \dd^2u \,
f_\e(u,D) \, \exp\Big[ - \frac1{2\e^2}D^2 - i \zeta D \Big] \;.
\ee
Here $f_\e(u,D)$ is the result of the path integral over all fields
except the zero-modes of the gauge field and $D$.
For $u \not\in \fM_\text{sing}$ and for $D$ whose
imaginary part is close enough to $0$ so that the real part of the term
$|D_{\mu}\phi|^2 + i \bphi D \phi$ is positive definite,  it has a limit as $\e\to 0$:
\bea
\label{hg}
f_\e(u,D) &\xrightarrow[\e\,\to\,0]{} \int \dd\lambda_0 \, \dd\bar\lambda_0\,
\bigg\langle \int \dd^2x\, \lambda \sum_i Q_i \bar\psi_i \phi_i  \int \dd^2x\, \bar\lambda \sum_i Q_i \psi_i \bar\phi_i \bigg\rangle_\text{free} \\
&\qquad=   h(\tau,z,u,D) \, g(\tau,z,u,D) \;,
\eea
where
\be
\label{eq:g}
g(\tau,z,u,D) = Z_\text{vect}(q,y)  \prod\nolimits_i Z_{\Phi,Q_i}(\tau,z,u,D)
\ee
is the one-loop determinant and
\be
h(\tau,z,u,D)=
c \sum_{i,n,m} \frac{Q_i^2}{\displaystyle
\left( \left| m + n \tau + Q_i u + \tfrac{R_i}2 z \right|^2+i Q_iD \right)
\left( m + n \bar\tau + Q_i \bar u + \tfrac{R_i}2 \bar z \right)} \;
\ee
arises from saturating the gaugino zero-modes.
The overall constant $c$ only depends on our normalization of the vector multiplet;
it can be fixed by comparing our final result to just one example in the limit $z=0$, and
it turns out to be $c = - i/\pi$.

\subsubsection{Deformation of the contour of $D$-integration}

To take the limit $\varepsilon \to 0$, it will be useful to deform the integral over $D$ away from the point $D=0$. In fact the contour can be deformed from the original $D \in \bR$, as long as the deformation does not hit poles of \eqref{hg}. Such poles are at
\be
D = \frac i{Q_i} \, \Big| m + n\tau + Q_i u + \frac{R_i}2 z \Big|^2 \;,
\ee
and they approach the real axis as $u$ approaches the singular points $u_* \in \fM_\text{sing}$. As we excised the regions $\Delta_\varepsilon$ in \eqref{int2}, the poles closest to the real axis have an imaginary part of order $\varepsilon^2$. So, let us define the contours $\Gamma_\pm$ to be $D \in \bR \pm i\delta$ with $0<\delta\ll \varepsilon^2$. We take $\Gamma_-$ for definiteness.

Now we use the following feature of \eqref{hg}:
\be
h(\tau,z,u,D) \, g(\tau,z,u,D)=  - \frac1{\pi D} \,
\frac{\partial}{\partial\overline{u}} g(\tau,z,u,D) \;,
\ee
to rewrite the elliptic genus as%
\footnote{Note that $\dd^2u \equiv \dd(\re u) \wedge \dd(\im u) = \frac i2 \dd u \wedge \dd\bar u$, and $\partial(\fM \setminus \Delta_\varepsilon) = - \partial\Delta_\varepsilon$.}
\be
Z_{T^2} = \lim_{\e,\varepsilon\to0}
\int_{\Gamma_-}\dd D\, \frac1{2\pi i D} \exp \Big[ - \frac1{2\e^2}D^2 - i \zeta D \Big]
\oint_{\partial \Delta_\varepsilon}\dd u\, g(u,D) \;.
\ee
Now we are interested in the $D$-plane poles of $g(u,D),$ which is defined in \eqref{eq:g}.  From the chiral multiplet contribution \eqref{eq:chiralproduct} we see that there are poles at $D =  i Q_i |u_*|^2.$  Since the contour of integration in the $u$-plane is restricted to $|u| = \epsilon_{*}$ we see that there are poles at $D = i Q_i\varepsilon_*^2.$
Let us look at one component of $\partial\Delta_{\varepsilon}$, that encircles
a singular point $u_*$. In addition to $D=0$,
there is a pole at $D = i Q_i\varepsilon_*^2$ for each $Q_i$ satisfying
(\ref{sing}) for $u=u_*$. On the same side of the real axis
there are other poles on the imaginary axis, while there is no pole on the other side of the line.
Recall that the signs of such $Q_i$'s are fixed:
they are positive if $u_* \in \fM_\text{sing}^+$ and
negative if $u_* \in \fM_\text{sing}^-$.
We decompose $\Delta_{\varepsilon}$ into two groups accordingly,
$\Delta^{(+)}_{\varepsilon}$ and $\Delta^{(-)}_{\varepsilon}$,
and separate the discussion of the integral into the two cases.

\begin{figure}[t]
\[
\def\scale{.48}
\begin{tikzpicture}[scale=\scale]
\node at (-12.5,0){
\begin{tikzpicture}[scale=\scale,>=stealth]
\draw[->] (-5,0) to (5,0);
\draw[->] (0,-5) to (0,5);
\draw (4,4) to (4,5);
\draw (4,4) to (5,4);
\node at (4.5,4.5) {$D$};
\node at (0,0) {$\times$};
\node at (0,2) {$\times$};
\node[anchor=east] at (-0.3,2) {$i Q_i \varepsilon^2_*$};
\draw[->] (0.5,2) to (0.5,1);
\node at (0,3) {$\times$};
\node at (0,3.5) {$\times$};
\draw[very thick] (-5,-1) to (5,-1);
\draw[->,very thick] (3,-1) to (3.1,-1);
\node[anchor=north] at (3,-1.3) {$\Gamma_-'$};
\end{tikzpicture}
};
\draw[dashed] (-6.5,-6) to (-6.5,6);
\node at (0,0) {
\begin{tikzpicture}[scale=\scale,>=stealth]
\draw[->] (-5,0) to (5,0);
\draw[->] (0,-5) to (0,5);
\draw (4,4) to (4,5);
\draw (4,4) to (5,4);
\node at (4.5,4.5) {$D$};
\node at (0,-0) {$\times$};
\node at (0,-2) {$\times$};
\node[anchor=east] at (-0.3,-2) {$i Q_i \varepsilon^2_*$};
\draw[->] (0.3,-2) to (0.3,-1.2);
\node at (0,-3) {$\times$};
\node at (0,-3.5) {$\times$};
\draw[very thick] (-5,-1) to (5,-1);
\draw[->,very thick] (3,-1) to (3.1,-1);
\node[anchor=north] at (3,-1.3) {$\Gamma_-$};
\node at (6,0) {$=$};
\end{tikzpicture}
};
\node at (11,0) {
\begin{tikzpicture}[scale=\scale,>=stealth]
\draw[->] (-5,0) to (5,0);
\draw[->] (0,-5) to (0,5);
\draw (4,4) to (4,5);
\draw (4,4) to (5,4);
\node at (4.5,4.5) {$D$};
\node at (0,-0) {$\times$};
\node at (0,-2) {$\times$};
\node[anchor=east] at (-0.3,-2) {$i Q_i \varepsilon^2_*$};
\draw[->] (0.3,-2) to (0.3,-1.2);
\node at (0,-3) {$\times$};
\node at (0,-3.5) {$\times$};
\draw[very thick] (-5,1) to (5,1);
\draw[->,very thick] (3,1) to (3.1,1);
\node[anchor=south] at (3,1.2) {$\Gamma_+'$};
\draw[very thick] (0,0) circle (.6) ;
\draw[very thick,->] (.5,.35) -- (.2,.6);
\node at (1.1,-.9) {$C_0$};
\end{tikzpicture}
};
\end{tikzpicture}
\]
\caption{Left: poles for $u_* \in \partial\Delta_\varepsilon^{(+)}$ and contour $\Gamma'_-$. Right: poles for $u_* \in \partial\Delta_\varepsilon^{(-)}$ and contour $\Gamma_-$, which is equivalent to the sum of $\Gamma'_+$ and $C_0$. \label{fig: poles}}
\end{figure}
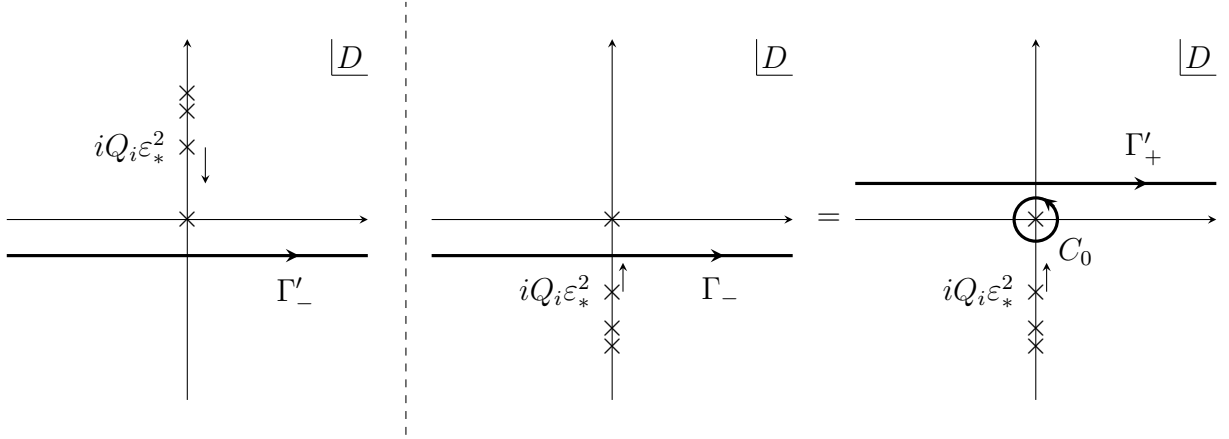

Let us first compute the contribution from the components of
$\partial\Delta^{(+)}_{\varepsilon}$, see figure~\ref{fig: poles} on the left.
There is no pole on the lower-half $D$-plane, therefore the contour $\Gamma_-$ can be deformed further away from $D = 0$ (relaxing the condition $\delta \ll \varepsilon^2$)
to an arbitrary position in the lower half-plane.
In particular we can take the $\varepsilon \to 0$ limit holding the new
contour $\Gamma'_-$ fixed. Since the integrand is continuous
and bounded as a function of $u$, while the integration region shrinks as $\varepsilon \to 0$, the integral vanishes in the limit. Thus we find:
\be
\label{vanish-}
\lim_{\e,\varepsilon\to0}
\int_{\Gamma'_-}\dd D\, \frac1{2\pi i D} \exp \Big[ - \frac1{2\e^2}D^2 - i \zeta D \Big]
\oint_{\partial \Delta_\varepsilon^{(+)}}\dd u\, g(u,D) = 0 \;.
\ee
Next consider the components of $\partial\Delta^{(-)}_\varepsilon$, see figure~\ref{fig: poles} on the right.
There are no poles in the upper half-plane but there are
infinitely many poles on the negative imaginary axis.
In each component, the closest poles to $D = 0$ are at $D = i Q_i \varepsilon_*^2$ for the corresponding set of $i$'s, and the contour $\Gamma_-$ goes between them.
In particular, the pole at $D = i Q_i\varepsilon_*^2$ approaches the pole at $D=0$
in the limit we are going to take.
No other poles approach the real line as $\varepsilon \to 0$.
Therefore we cannot take the limit $\varepsilon \to 0$ while holding the contour fixed.
To avoid this complication, we decompose the contour as
\be
\Gamma_- = \Gamma'_+ + C_0
\ee
where $C_0$ is a circle of radius smaller than $\varepsilon_*^2$
that goes around $D = 0$ counter-clockwise.
For the same reason as in (\ref{vanish-}) we have
\be
\label{vanish+}
\lim_{\e,\varepsilon\to0}
\int_{\Gamma'_+}\dd D\, \frac1{2\pi i D} \exp \Big[ - \frac1{2\e^2}D^2 - i \zeta D \Big]
\oint_{\partial \Delta_\varepsilon^{(-)}}\dd u\, g(u,D) = 0 \;,
\ee
and we are left with
\bea
\label{u1formula1}
Z_{T^2} &= \lim_{\e, \varepsilon\to0} \int_{C_0}\dd D\, \frac1{2\pi i D} \exp \Big[ - \frac1{2\e^2}D^2 - i \zeta D \Big]
\oint_{\partial \Delta_\varepsilon^{(-)}}\dd u\, g(u,D) \\
&= \lim_{\e,\varepsilon\to0} \oint_{\partial \Delta^{(-)}_{\varepsilon}}\dd u\,
g(u,0) \\
&= \oint_{\partial \Delta^{(-)}_{\varepsilon}}{\dd u}\, g(u,0)
\eea
which corresponds to the formula \eqref{u1formula neg} we wanted to derive.

If we instead choose $\Gamma = \Gamma_+$ then by repeating the same argument, mutatis mutandis, we find:
\bea
Z_{T^2} &= \lim_{\e,\varepsilon\to0} \int_{-C_0}\dd D\, \frac1{2\pi i D} \exp \Big[ - \frac1{2\e^2}D^2 - i \zeta D \Big]
\oint_{\partial \Delta_\varepsilon^{(+)}}\dd u\, g(u,D) \\
&= - \lim_{\e,\varepsilon\to0} \oint_{\partial \Delta^{(+)}_{\varepsilon}}\dd u\,
g(u,0) \\
&= - \oint_{\partial \Delta^{(+)}_{\varepsilon}}{\dd u}\, g(u,0) \;.
\eea
This is precisely our formula \eqref{u1formula}.
The two answers for $Z_{T^2}$, \eqref{u1formula} and \eqref{u1formula neg}, are indeed the same since
$g(u,0)$ is holomorphic on $\fM \setminus \Delta_{\varepsilon}$. In particular the choice of a ``displacement vector'' $\pm\delta$ of the contour of $D$ in the complex plane is necessary to perform the computation, but it does not affect the final answer.

\section{Examples}
\label{sec: examples}

We will finally present some illustrative examples, with each one elucidating a different aspect.

\subsection{The quintic}
\label{sec: the quintic}

Let us start with a classic example, the quintic Calabi-Yau, which is an Abelian rank-one theory that flows to a conformal fixed point.
Consider a $\UU(1)$ theory with one chiral multiplet $P$ of charge $-5$ and five chiral multiplets $X_i$ of charge $1$, and a superpotential $W=Pf(X_1,\ldots,X_5)$ where $f$ is a quintic polynomial.%
\footnote{We assign R-charge 2 to $P$ and 0 to $X_i$, although this can be changed by an unphysical mixing with the gauge symmetry.}
The formula (\ref{u1formula}) applied to $\fM_\text{sing}^-$, in other words summing over the residues in $\fM_\text{sing}^-$, gives the elliptic genus as
\bea
\label{quinticLG}
Z_{T^2}(\tau,z) &= \frac{i\eta(q)^3}{\theta_1(q,y^{-1})} \; \sum_{k,l=0}^4 \; \oint_{u = (z+k+l\tau)/5} \hspace{-1em} \dd u\, \frac{\theta_1(q,x^{-5}) }{\theta_1(q, y \, x^{-5})} \left(\frac{\theta_1(q,y^{-1} \, x) }{\theta_1(q, x)}\right)^5 \\
&=  \frac15 \sum_{k,l=0}^4  y^{-l} \left(\frac{\theta_1(q,y^{-4/5} e^{2\pi i (k+l\tau)/5}) }{\theta_1(q, y^{1/5} e^{2\pi i (k+l\tau)/5 })}\right)^5 \;.
\eea
The second equality follows from
$$
\oint_{u=0} \dd u \, \frac{\theta_1 ( \tau | z + k + l\tau)}{\theta_1(\tau | u - k - l\tau)} = y^{-l} \, \frac{\theta_1(q,y)}{2\pi \, \eta(q)^3} \;,
$$
see also Appendix \ref{app: theta}. If we instead apply formula (\ref{u1formula}) to $\fM_\text{sing}^+$, the expression for the elliptic genus is
\be
\label{quinticgeom}
Z_{T^2}(\tau,z) = - \frac{i\eta(q)^3}{\theta_1(q,y^{-1})} \oint_{u=0} \dd u\, \frac{\theta_1(q,x^{-5}) }{\theta_1(q, y\,x^{-5})} \left(\frac{\theta_1(q,y^{-1} \, x) }{\theta_1(q, x)}\right)^5 \;.
\ee

The final expression in \eqref{quinticLG} is the standard expression of the elliptic genus of the Landau-Ginzburg orbifold with $W = f(X_1,\ldots,X_5)$ with five fields of R-charges $2/5$ under the diagonal $\bZ_5$ action. Using
$$
\lim_{\tau \to i\infty} \frac{\theta_1(q,y^{-4/5} e^{2\pi i (k+l\tau)/5}) }{\theta_1(q, y^{1/5} e^{2\pi i (k+l\tau)/5 })} = \begin{cases} -y^{3/10} e^{-2\pi i k/5} \frac{1-y^{-4/5} e^{2\pi i k/5}}{1-y^{-1/5} e^{-2\pi i k/5}} &\text{for } l=0 \\ y^{1/2} & \text{for } l = 1,2,3,4, \end{cases}
$$
we can obtain the $\chi_y$ genus in the $\tau \to i\infty$ limit:
$$
\lim_{\tau \to i\infty} Z_{T^2}(\tau,z) = -100 \, y^{1/2} - 100 \, y^{-1/2} \;,
$$
and the Witten index equal to the Euler number of the quintic: $\chi = Z_{T^2}(\tau,0) = -200$.
On the other hand, the right hand side of \eqref{quinticgeom} is the standard geometric expression of the elliptic genus of the quintic hypersurface $X$ in $\CP^4$. Indeed, the elliptic genus $Z_X(\tau,z)$ is mathematically defined as the integral of a characteristic class:
\be
Z_\text{geom}(\tau,z) = - \frac{i\eta(q)^3}{\theta_1(q,y^{-1})} \int_X  \prod_a \frac{x_a \, \theta_1(q,y^{-1} \, e^{x_a} )}{\theta_1(q, e^{x_a})}
\ee
where $x_a$ are the Chern roots of the tangent bundle of $X$.
In this case $X$ is given by the divisor $5H$, where $H$ is the hyperplane class of $\CP^4$. We thus have
\be
Z_\text{geom}(\tau,z) = -\frac{i\eta(q)^3}{\theta_1(q,y^{-1})}
\int_{\CP^4} H^4 \, \frac{\theta_1(q,e^{-5H})}{\theta_1(q, y \, e^{-5H})} \,
\left( \frac{\theta_1(q,y^{-1} \, e^{H} )}{\theta_1(q, e^H)} \right)^5
\ee
which is literally equal to \eqref{quinticgeom} once we use $\int_{\CP^4} H^4 f(H) = \oint_{u=0} \dd u\, f( 2\pi i u)$. For more details and generalizations see Appendix \ref{app: math}.

\subsection{$\WW\PP_{1,1,2,2,2}^4[8]$ }
\label{sec: the wp}

Our next example is a gauged linear sigma model that describes a Calabi-Yau hypersurface with an orbifold singularity in a weighted projective space. Specifically, we consider a degree 8 hypersurface in $\WW\PP^4_{1,1,2,2,2}$.  We will see below that our formula automatically takes care of the orbifold singularities.

This geometry is realized by a $\UU(1)$ theory with one chiral multiplet $P$ of charge $-8$, two  chiral multiplets $X_i$ of charge $1$, three chiral multiplets $Y_j$ of charge $2$, and a superpotential $W=Pf(X_1,X_2,Y_1,Y_2,Y_3),$ where $f$ is a charge-8 polynomial.
We assign R-charge 2 to $P$ and zero R-charge to $X_i$ and $Y_j$.
The formula (\ref{u1formula}) applied to $\fM_\text{sing}^-$ gives the elliptic genus as
\begin{align}
Z_{T^2}(\tau,z) &= \frac{i\eta(q)^3}{\theta_1(q,y^{-1})} \; \sum_{k,l=0}^7 \; \oint_{u = (z+k+l\tau)/8} \hspace{-1em} \dd u\, \frac{\theta_1(q,x^{-8}) }{\theta_1(q, y \, x^{-8})}  \left(\frac{\theta_1(q,y^{-1} \, x) }{\theta_1(q, x)}\right)^2  \left(\frac{\theta_1(q,y^{-1} \, x^{2}) }{\theta_1(q, x^2)}\right)^3 \nn \\
&=  \frac18 \sum_{k,l=0}^7  y^{-l} \left(\frac{\theta_1(q,y^{-7/8} e^{ 2 \pi i (k+l\tau)/8 ) } }{\theta_1(q, y^{1/8} e^{2\pi i (k+l\tau)/8 })}\right)^2
\left(\frac{\theta_1(q,y^{-3/4} e^{2\pi i (k+l\tau)/4 ) } }{\theta_1(q, y^{1/4} e^{2\pi i (k+l\tau)/4 })}\right)^3 \;.
\end{align}
This is the expression of the elliptic genus in the Landau-Ginburg phase, which is a $\bZ_8$ orbifold theory. Instead using $\fM_\text{sing}^+$ to compute the elliptic genus we find
\be
\label{WP genus geometric}
Z_{T^2}(\tau,z) = - \frac{i\eta(q)^3}{\theta_1(q,y^{-1})}
\sum_{k,l=0}^1 \oint_{u=(k+l\tau)/2}  \hspace{-1em} \dd u\, \frac{\theta_1(q,x^{-8}) }{\theta_1(q, y \, x^{-8})}
\left(\frac{\theta_1(q,y^{-1} \, x) }{\theta_1(q, x)}\right)^2
\left(\frac{\theta_1(q,y^{-1} \, x^{2}) }{\theta_1(q, x^2)}\right)^3 .
\ee
The two expressions agree, and in the limit $\tau\to i\infty$ they give
$$
\lim_{\tau \to i\infty} Z_{T^2}(\tau,z) = - 84 \, y^{1/2} - 84 \, y^{-1/2} \;.
$$
Taking the limit of (\ref{WP genus geometric}), the pole at $u=0$ contributes $-81(y^{1/2}+y^{-1/2})$ while the other three poles contribute $-(y^{1/2}+y^{-1/2})$ each. We correctly find the Witten index  of the model to be $Z_{T^2}(\tau,0) = -168$.

Geometrically, the degree-8 hypersurface in $\WW\PP_{1,1,2,2,2}^4$ has Euler number $-162$,
but it also has a genus 3 curve of $A_1$ singularities: blowing it up  introduces an additional Euler number $-6$, giving the Euler number of the resolved smooth Calabi-Yau space to be $-168$.
We find that the Witten index of the theory computed with our method correctly accounts for the contribution from the orbifold singularity.

\subsection{$\CC\PP^{N-1}$}
\label{sec: CPN}

Our third example is a massive Abelian rank-one theory which instead of flowing to a fixed point develops a mass gap.

Consider a $\UU(1)$ theory with $N$ chiral multiplets $\Phi_i$ of gauge charge $+1$. The left-moving R-symmetry is anomalous: indeed the one-loop determinant $Z_\text{1-loop}(\tau,z,u)$ has monodromies on the torus:
\be
Z_\text{1-loop}(\tau, z, u+a + b\tau) = e^{2\pi i b z N} Z_\text{1-loop}(\tau,z,u) \qquad\qquad\text{for } a,b\in\bZ \;.
\ee
Single-valuedness of $Z_\text{1-loop}(\tau,z,u)$ requires $z \in \bZ/N$.

Applying the formula (\ref{u1formula}) to $\fM_\text{sing}^-$ we immediately get that $Z_{T^2}(\tau,z) = 0$, because $\fM_\text{sing}^-$ is empty. However we cannot trust this result for $z=0$ because in that case the integrand $Z_\text{1-loop}$ is not well-defined. To proceed, we introduce an extra chiral multiplet $P$ of gauge charge $Q_P = -N$: now the theory has unbroken left-moving R-symmetry, we can consider arbitrary $z$ and analytically continue to $z=0$. We assign R-charge 1 to $P$ in such a way that $P$ does not contribute to the index because the bosonic and fermic contributions cancel out and hence $Z_P = 1$. Of course the theory with $P$ is not the same as the theory without it. However at $z=0$ we can turn on a twisted mass%
\footnote{The twisted mass has R-charge 2, so it is forbidden by a non-trivial R-symmetry holonomy $z$.}
for $P$ and the extra chiral multiplet decouples.

It is also convenient to introduce generic flavor holonomies%
\footnote{The flavor symmetry is $SU(N)$, therefore the flavor holonomies should satisfy $\sum_k \xi_k = 0$. It is easier to allow generic $\xi_k$, keeping in mind that their sum can be absorbed by a shift of the integration variable $u$. On a flat manifold as $T^2$ this does not produce any effect at all.}
$-\xi_{k=1, \cdots,N}$. The elliptic genus (\ref{u1formula}) is then given by
\be
Z_{T^2}(\tau,z, -\xi_k) = \sum_{u_j \,\in\, \fM_\text{sing}^+} \oint_{u=u_j} \dd u\, \frac{i \eta(q)^3}{\theta_1(q,y)} \prod_{k=1}^N \frac{\theta_1(\tau | -z+u-\xi_k)}{\theta_1(\tau|u- \xi_k)} \;.
\ee
There are $N$ simple poles at $u=\xi_k$, and the result is
\be
Z_{T^2}(\tau, z, -\xi_k) =  \sum_{k=1}^N \prod_{j \, (\neq k)} \frac{\theta_1(\tau|-z + \xi_k - \xi_j)}{\theta_1(\tau| \xi_k - \xi_j)} \;.
\ee

For $z=0$ we find $Z_{T^2} =  N$, which is the Witten index of $\CP^{N-1}$.
 The limit $\tau \to i\infty$, \ie{} $q\to 0$, instead gives the $\chi_y$ genus of $\CP^{N-1}.$  Using
$$
\frac{\theta_1(\tau|a)}{\theta_1(\tau|b)} \xrightarrow[q\to0]{} \frac{e^{i\pi a} - e^{-i\pi a}}{e^{i\pi b} - e^{-i\pi b}} \, \big( 1 + \cO(q) \big)
$$
we find:
\be
\label{chi y genus P^N}
Z_{T^2}(i\infty, z, -\xi_k ) = y^{- \frac{N-1}2} \sum\nolimits_{j=0}^{N-1} y^j \;.
\ee
Notice that there is no dependence on $\xi_j$. This is as it should be, since the harmonic forms representing the cohomology classes are invariant under the isometry.

\subsection{$\SU(2)$ with fundamentals}
\label{sec: SU2}

Let us next consider an $\cN=(2,2)$ $\SU(2)$ gauge theory with $N$ fundamental
chiral multiplets.
Here we evaluate the elliptic genus, the $\chi_y$ genus
(\ie{} its $q \to 0$ limit) and the Witten index using our formalism.
The elliptic genus is
\begin{multline}
Z_{T^2}(\tau, z, \xi) = - \frac12 \sum_{u_* \,\in\, \fM^+_\text{sing}}
\frac{i \eta(q)^3}{\theta_1(\tau|-z)}
\oint_{u_*}\dd u\, \frac{\theta_1(\tau|2u)}{\theta_1(\tau|-z+2u)} \,
\frac{\theta_1(\tau|-2u)}{\theta_1(\tau| -z - 2u)} \\
\prod_{j=1}^N \frac{\theta_1\big( \tau \big|
(\frac R2-1)z + u + \xi_j \big)}{\theta_1 \big( \tau \big|
\frac R2 z + u + \xi_j \big)} \, \frac{\theta_1 \big( \tau \big|
(\frac R2-1)z - u + \xi_j \big)}{\theta_1\big( \tau \big|
\frac R2 z - u + \xi_j \big)} \;.
\end{multline}
We have assigned R-charge $R$ to the fundamentals,
and we have introduced flavor holonomies $x_j=e^{2\pi i \xi_j}$
with $j=1,\dots,N$ in the Cartan of $\UU(N)$.
The set of poles is read off from the denominators.
The set $\fM^+_\text{sing}$ is
\be
\fM^+_\text{sing}=\Big\{  \frac z2 \,,\,
\frac{z+1}2 \,,\,  \frac{z+\tau}2 \,,\,  \frac{z+\tau+1}2
\,,\,  -\xi_j - \frac R2 z \Big\} \;.
\ee

Let us consider the theory with vanishing superpotential, in which case
the correct R-charge in the infra-red limit is $R=0$ and we can turn on
an arbitrary flavor holonomy.
Evaluating the various contributions
(the formula \eqref{boo} will be useful), we find:
\begin{multline}
Z_{T^2}(\tau,z,\xi)
= - \frac14 \, \frac{\theta_1(\tau|z)}{\theta_1(\tau|2z)}
\sum_{a,b=0}^1 y^{-bN} \prod_{j=1}^N \frac{\theta_1(\tau| - \tfrac32 z
+ \frac{a+b\tau}2 + \xi_j )}{\theta_1( \tau|
\frac z2 + \frac{a+b\tau}2 + \xi_j)} \\
+ \frac12 \sum_{j=1}^N \frac{\theta_1(\tau| 2\xi_j)}{\theta_1(\tau| 2\xi_j + z)}
\prod_{k\, (\neq j)}^N \frac{\theta_1(\tau|-\xi_j + \xi_k - z) \,
\theta_1 (\tau| -\xi_j - \xi_k + z)}{\theta_1(\tau|-\xi_j + \xi_k) \;
\theta_1(\tau| -\xi_j - \xi_k)} \;.
\end{multline}
The theory has a non-compact moduli space of classical vacua:
both a Higgs branch (for $N\geq 2$) and a Coulomb branch.
To control their contributions,
let us first take the limit $q\to 0$ for generic $x_i$'s and $y$, and then
send the flavor holonomies to infinity: $x_i \to 0$. We find
\be
\chi_y \equiv \lim_{x\to0} \lim_{q \to 0} Z_{T^2} =
y^{3/2}\frac{1+y+y^2+\cdots+y^{N-2}}{1+y}.
 \;\label{chiy}
\ee
The limit $x_i\to0$ has killed a possible Higgs branch completely,
but has no effect on the Coulomb branch.
In the quantum theory, the Coulomb branch is lifted for odd $N$
but remains unlifted for even $N$ \cite{Hori:2006dk}.

When $N$ is odd, the expression above can be simplified to
\be
\chi_y = y^{3/2}+y^{7/2}+\cdots+y^{N-3/2} \;.
\ee
This has no divergence whatsoever.
If we set $y=1$, we obtain the Witten index $(N-1)/2$,
which agrees with the Witten index of the theory deformed by a
generic twisted mass, computed in \cite{Hori:2006dk}.
This agreement is expected for odd $N$:
the present computation and
the computation in \cite{Hori:2006dk} can be continuously connected
by a path in the space of the twisted masses and the flavor holonomy
on the torus. For a generic choice of path,
the Witten index is well-defined along the way
and hence it is constant \cite{Witten:1982df}.
Therefore the initial and the final values must agree.

As another application, let us compute the $\chi_y$ genus
of the theory with $N=3$ for generic $x_i$'s:
\begin{equation}
\chi_y = \frac{1}{y^{3/2}}
\frac{y- x_1 x_2}{1-x_1 x_2}
\frac{y- x_2 x_3}{1-x_2 x_3}
\frac{y- x_3 x_1}{1-x_3 x_1}.
\end{equation}
This is the same as the $\chi_y$ genus of three free fields
which carry the same charges as the baryons.
This is consistent with the claim in  \cite{Hori:2006dk}
that the $SU(2)$ theory with $3$ fundamentals
flows to the free theory of the three baryons.

For $N$ even, the $\chi_y$ genus \eqref{chiy} has a divergence at  $y=-1$,
caused by the contribution of a non-compact Coulomb branch
parameterized by $\tr \Sigma^2$, where $\Sigma$ is the adjoint twisted
chiral multiplet constructed from the vector multiplet.
Its contribution to the $\chi_y$ genus is, from \eqref{twisted chiral},
\be
\frac{y^{1/2}}{1+y} \;.
\ee
This has indeed the same kind of divergence as \eqref{chiy} at $y=-1$.

\subsection{$\OO(2)$ with fundamentals}
\label{sec:O2}

As the last example with $\cN=(2,2)$ supersymmetry,
we consider  an $\OO(2)$ gauge theory with $N$ chiral multiplets
in the fundamental doublet.
We include this example to show how the computation works
in a theory with non-connected gauge group.
$\OO(2)$ can be regarded as $\UU(1)\rtimes \ZZ_2$, where the generator
$\gamma\in \ZZ_2$ acts on $\UU(1)$ as the inversion, and a doublet consists
of two fields $\phi_1,\phi_2$ of $\UU(1)$ charge $1,-1$ which are exchanged by
the element $(1,\gamma)\in \UU(1)\rtimes \ZZ_2$.

The moduli space of flat $\OO(2)$ connections on the torus consists of
seven components: one is the space of $u\in \CC/(\ZZ+\tau\ZZ)$ modulo
$u\equiv -u$, and the others are six points represented by
the commuting pairs of holonomies
$\big( (1,\gamma),(\pm 1, 1) \big)$, $\big( (\pm 1,1),(1,\gamma) \big)$,
$\big( (\pm 1, \gamma),(1,\gamma) \big)$. The computation on the first component is
as before. For the computation at the six discrete holonomies,
take the combinations $\phi_{\pm}=\phi_1\pm \phi_2$.
Both of them are odd under the element
$(-1,1)$ while $\phi_+$ ({\it resp}.~$\phi_-$) is even ({\it resp}.~odd)
under $(1,\gamma)$.
The $\UU(1)$ gauge multiplet is even under $(-1,1)$ and odd under
$(1,\gamma)$.
The theory carries a discrete data $(\epsilon,\theta)\in
\{+1,-1\}\times \{0,\pi\}$
which specifies a weight for the sum over components. The elliptic genus is given by
\bea
& Z^{\epsilon,\theta}_{T^2}(\tau,z,\xi) = \\
&\qquad - \frac12 \sum_{j=1}^N \frac{i \eta(q)^3}{\theta_1(\tau|-z)}
\oint_{-\xi_j-{R\over 2}z} \hspace{-.5em} \dd u\,
\prod_{j=1}^N \frac{ \theta_1 \big(\tau \big| (\frac R2-1)z + u + \xi_j \big)}{ \theta_1 \big(\tau \big| \frac R2 z + u + \xi_j \big)} \,
\frac{ \theta_1 \big(\tau \big| (\frac R2-1)z - u + \xi_j \big) }{ \theta_1 \big( \tau \big| \frac R2 z - u + \xi_j \big)} \\
&\qquad - \frac14 \sum_a \epsilon_a \, e^{i\theta_a} \,
\frac{\theta_1(\tau|a_{\rm v}) }{ \theta_1(\tau|z-a_{\rm v})}
\prod_{j=1}^N \frac{\theta_1 \big( \tau \big| ({R\over 2}-1)z+a_+ + \xi_j \big) }{ \theta_1 \big(\tau \big| {R\over 2}z+a_+ + \xi_j \big)}
\, \frac{ \theta_1 \big( \tau \big| ({R\over 2}-1)z+a_- + \xi_j \big) }{ \theta_1 \big(\tau \big|{R\over 2}z+a_- + \xi_j \big)} \;.
\eea
The numbers $(a_{\rm v},a_+,a_-)$ and
$(\epsilon_a,\theta_a)$ for the discrete holonomy $a$ are given by:
\be
\begin{array}{c|ccc}
\phantom{\Big|} a & \big( (1,\gamma),(1,1) \big) & \big( (1,\gamma),(-1,1) \big) & \big( (1,1),(1,\gamma) \big) \\
\hline
\phantom{\Big|} (a_{\rm v},a_+,a_-) &({1\over 2},0,{1\over 2}) & ({1\over 2},-{\tau\over 2},{1+\tau\over 2}) & ({\tau\over 2},0,{\tau\over 2}) \\
(\epsilon_a,\theta_a) & (1,1)&(1,\theta) & (\epsilon,1) \\[1.5em]
\hline\hline
\phantom{\Big|} a & \big( (-1,1),(1,\gamma) \big) & \big( (1,\gamma),(1,\gamma) \big) & \big( (-1,\gamma),(1,\gamma) \big) \\
\hline
\phantom{\Big|} (a_{\rm v},a_+,a_-) & ({\tau\over 2},-{1\over 2},{1+\tau\over 2})&
({1+\tau\over 2},0,{1+\tau\over 2})&
({1+\tau\over 2},-{1\over 2},{\tau\over 2})\\
(\epsilon_a,\theta_a)  &(\epsilon,\theta)&
(1,1)&(1,\theta)
\end{array}
\ee
Here $\epsilon=\pm 1$ corresponds to a choice of the action of $\gamma$
on the untwisted RR sector \cite{Dixon:1988ac,Intriligator:1990ua},
when the theory is viewed as a $\ZZ_2$ orbifold of the $\UU(1)$ gauge theory.
$\theta\in\{0,\pi\}$ is the discrete theta angle.

Let us examine the formula for the theory with no superpotential.
In this case, the right value of the R-charge in the infrared limit is
$R=0$ and we can turn on any flavor holonomy.
The theory is plagued by a non-compact Coulomb branch, unless
the discrete theta angle is trivial ($\theta=0$)
for odd $N$ or non-trivial ($\theta=\pi$) for even $N$
\cite{Hori:2011pd}.
It is easy to evaluate the limit $q\to 0$ first, and then $y\to 1$, leaving
$\xi_i$ generic:
\be
\label{O2index}
\chi_{y=1} = \frac N2  + \frac{1+ e^{i\theta}}4 + \epsilon \frac{1+ e^{i\theta}}4  + \frac{1+ e^{i\theta}}4
= \begin{cases} \dfrac{N+3}2 & \text{for odd $N$, $\theta=0$, $\epsilon=+1$}, \\[0.4cm]
\dfrac{N+1}2 & \text{for odd $N$, $\theta=0$, $\epsilon=-1$}, \\[0.4cm]
\dfrac N2  & \text{for even $N$, $\theta=\pi$, $\epsilon=\pm 1$}.
\end{cases}
\ee
These are the cases where the theory is regular by the right choice
of theta angle. For the non-regular values, it would give a non-sensical
non-integer.
The result (\ref{O2index}) agrees with the result of \cite{Hori:2011pd}
for the Witten index of the theory with generic twisted masses.
The reason of the agreement is the same as in the $SU(2)$ case.

Let us look at the $\chi_y$ genus for a general $y$
in the regular theory with $N=1$:
\be
\chi_y=\left({1\over 2}+{1\over 2}+{\epsilon\over 2}+{1\over 2}\right)
{1\over y^{1/2}}{y-x^2\over 1-x^2}
={1\over y^{1/2}}{y-x^2\over 1-x^2}
\times\left\{
\begin{array}{ll}
2,&\epsilon=+1,\\
1,&\epsilon=-1.
\end{array}\right.
\ee
This is the same as the $\chi_y$ genus of a single free field
which carries the same charges as the meson $\phi_1\phi_2$.
This is consistent with the claim in \cite{Hori:2011pd}
that the regular theory flows in the infrared limit to
one or two copies of the free theory of the meson,
where the number of copies depends on the choice of $\epsilon$.

\subsection{Distler-Kachru models}
\label{sec:DK}

Finally let us discuss  the elliptic genus of an $\cN=(0,2)$ gauged linear sigma model \cite{Distler:1993mk} which in the IR flows to a non-linear sigma model whose target is a Calabi-Yau $X$ with a stable holomorphic vector bundle $\cE \rightarrow X$.

Similar to our previous $\cN=(2,2)$ examples, we can engineer a Calabi-Yau $N$-fold hypersurface $X$ in weighted projective space $\WW\PP^{N+1}_{\{q_i\}}$ using a $\UU(1)$ gauge theory.  We take $N+2$ chiral multiplets $\Phi_i$ with gauge charges $q_i$ to construct the ambient weighted projective space.  We add a Fermi superfield $\Sigma$ with charge $-d$ and a superpotential term $J_\Sigma = W(\Phi)$, where $W(\phi)$ is a homogeneous polynomial of degree $d$, to cut out the hypersurface $W(\phi) = 0$.
A rank-$\tilde r$ holomorphic bundle $\cE$ on $X$, defined through the exact sequence
\be
0 \,\to\, \cE \,\to\, \bigoplus_{a=1}^{\tilde r+1} \cO(n_a) \,\xrightarrow{\oplus F_a}\, \cO(m) \,\to\, 0
\ee
where $F_a(\phi)$ are homogeneous polynomials of degree $m-n_a$, can be engineered by $\tilde r + 1$ Fermi superfields $\Lambda^a$ with charges $n_a$, a chiral superfield $P$ with charge $-m$ and superpotential terms $J_a = P F_a(\Phi)$.

The condition for cancellation of the gauge anomaly is
\be
\sum\nolimits_i q_i^2 + d^2 - \sum\nolimits_a n_a^2 - m^2 = 0 \;,
\ee
which corresponds to the anomaly-cancellation condition $c_2(T) = c_2(\cE)$ in the sigma model ($T$ is the tangent bundle to $X$). These models have a $U(1)_L$ flavor symmetry giving charges $(0,0,-1,1)$ to $(\Phi_i, \Sigma, \Lambda^a, P)$ respectively, and a $U(1)_R$ right-moving R-symmetry. The (global-gauge) anomaly cancellation conditions for these two symmetries are
\be
\label{eq:anom}
\sum\nolimits_a n_a - m = 0  \;,\qquad\qquad\qquad \sum\nolimits_i q_i - d = 0
\ee
and correspond to $c_1(\cE) = 0$ and $c_1(T) = 0$ in the sigma model.
When the Calabi-Yau target $X$ is smooth, twice the net number of generations---or equivalently the Witten index---equals
\be
c_3(\cE) = \frac13 \left(m^3 - \sum\nolimits_a n_a^3 \right) J^3 \;.
\ee
Finally, the gravitational anomaly is $k = N - \tilde r$.

\paragraph{$\boldsymbol{Y_{W\, 4;10}}$.} As a concrete example, consider the case of a CY$_3$ hypersurface in $\WW\PP_{1,1,1,2,5}^4$ with a rank-4 holomorphic vector bundle $\cE$ specified by $n_a = (1,1,1,1,7)$, which was discussed in \cite{Distler:1993mk}.%
\footnote{Concretely, the model has six chiral multiplets $\Phi_i,\Sigma$ with charges $(1,1,1,2,5,-10)$, six Fermi multiplets $\Lambda^a, P$ with charges $(1,1,1,1,7,-11)$, and an $\cN=(0,2)$ vector multiplet.}
Our formula for the $\cN=(0,2)$ elliptic genus is%
\footnote{We thank Sarah Harrison, Shamit Kachru and Natalie Paquette for pointing out an imprecision in a previous version of this section.}
\be
Z_{T^2}(\tau,z) = - 
\eta(q)^2 
\oint_{u=0} \dd u\, \frac{\theta_1(q, y^{-1}x)^4\, \theta_1(q, y^{-1} x^7) \, \theta_1(q, x^{-10})}{\theta_1(q,x)^3 \, \theta_1(q,x^2) \, \theta_1(q,x^5) \, \theta_1(q, y\, x^{-11})}
\ee
since there is a single positive pole. Note that $y$ is here the fugacity for the flavor symmetry $U(1)_L$. In the $q\to 0$ limit we get
\be
Z_{T^2}(\tau,z) = q^\frac1{12} \, \frac{1-y^2}y \, \chi(E) + \cO(q^\frac{13}{12}) \qquad\qquad\text{with}\qquad \chi(E) = 164
\ee
in agreement with \cite{Kawai:1994np}.

\section*{Acknowledgments}
The authors thank M.~Romo for collaboration at an early stage of this work
and N.~Bobev for helpful comments on an early version of the manuscript.
FB's work is supported in part by DOE grant DE-FG02-92ER-40697.
KH's work is supported in part by JSPS Grant-in-Aid for Scientific Research No. 21340109.
YT's work is supported in part by JSPS Grant-in-Aid for Scientific Research No. 25870159.
RE, KH, YT are also supported in part by WPI Initiative, MEXT, Japan at IPMU, the University of Tokyo.

\appendix

\section{Geometrical formula for the elliptic genus}
\label{app: math}

For a vector bundle $V$ with Chern roots $x_i$, let us define a characteristic class $\varphi(V)$ by
\be
\varphi(V)=\prod_i \frac{x_i}{f(x_i)}
\ee
where $f(x)$ is a formal power series in $x$ such that $f(0)=0$ and $f'(0)$ is finite. The reason for this definition will be clear later.
A generalized genus, in the sense of Hirzebruch, of an almost complex manifold $X$ is then
\be
\varphi(X)=\int_X \varphi(T_\bC X) \;.
\ee
For a readable introduction on the elliptic genus and other genera, see \cite{Hirzebruch}.

Our aim in this section is to find the formula for $\varphi(X)$ when $X$ is a complete intersection in  a K\"ahler quotient $M=V/\!/G$, where $G$ is a compact group and $V$ is a representation of $G$, by generalizing the argument in \cite{MaZhou,GuoZhou}.
Given a representation $R$ of $G$, we can construct a vector bundle $[R]$ on $M=V/\!/G$ whose fiber at a point is $R$. Let us say that $X$ is given by the common zeros of sections of a vector bundle $[E]$, where $E$ is a representation of $G$.

The elliptic genus $\varphi_\text{Ell}[X](q,y)$ is then given (see \eg{} \cite{Kawai:1993jk}) by choosing
\be
\label{mathelliptic}
f_{q,y}(x) =  y^{1/2} \prod_{n \geq 1} \frac{(1 - q^{n-1}e^{-x}) (1-q^n e^x)}{(1 - y q^{n-1} e^{-x})(1 - y^{-1} q^n e^x)} = \frac{\theta_1(\tau | \frac{x}{2\pi i})}{ \theta_1(\tau | \frac{x}{2\pi i}-z)} \;,
\ee
where $q=e^{2\pi i\tau}$ and $y=e^{2\pi i z}$, but the argument in this section applies to any genus.

First, we use the adjunction formula
\be
T_\bC M|_X = T_\bC X \oplus [E]|_X
\ee
to write
\be
\varphi(X)=\int_X \varphi(T_\bC X)=\int_X \frac{\varphi(T_\bC M)}{\varphi([E])} = \int_M  \varphi(T_\bC M) \, \frac{e([E])}{\varphi([E])} \;,
\ee
where we used the Euler class $e(\cdots)$ to express the integral over $X$ as an integral over $M$.
We note that $T_\bC M\oplus [\fg] =[V]$, where $\fg$ is the complexified Lie algebra of $G$. Therefore
\be
\label{fubar}
\varphi(X) = \int_M \frac{\varphi([V])}{\varphi([\fg])} \, \frac{e([E])}{\varphi([E])} \;.
\ee

The residue formula of Jeffrey and Kirwan \cite{JeffreyKirwan,Martens}---which originated from a conjecture by Witten \cite{Witten:1992xu}---in the case of a K\"ahler quotient of a vector space,
is the general statement that
\be
\int_{V/\!/G} c([R]) = \frac{1}{|W|}\oint \prod_{i=1}^{\rank G}  \frac{dz_i}{2\pi i} \, \frac{\prod_{\alpha: \text{ root of }\fg} \alpha(z)}{\prod_{v \in V} v(z) } \, \prod_{w \in E} \big( 1+w(z) \big)
\ee
where $z=(z_1,\ldots,z_{\rank G})$ takes values in the complexified Cartan subalgebra of $G$, $v$ and $w$ run over the weight vectors of the representations $V$ and $R$ respectively, so that $w(z)$ is the evaluation of a weight on an element of the Cartan subalgebra.  $|W|$ is the order of the Weyl group of $G$.
The repeated residue integral is taken around $(z_1,\ldots,z_r)=0$ on a particular cycle: this is called the Jeffrey-Kirwan residue. When all weight vectors $v$ lie on a half-space and the ambient space $V/\!/G$ is compact, this cycle is given as a sum of $T^r$'s around all unordered choices of $r$ hyperplanes $v_{i_1}(z)=0$, $v_{i_2}(z)=0$, \ldots, $v_{i_r}(z)=0$, as explained in \cite{SzenesVergne}.

In the following, we treat  a representation $R$ as a vector bundle whose Chern roots are $w(z)$.
Then we have $\prod_{v \in V} v(z)=e(V)$, for example.

Applying the residue formula to \eqref{fubar}, we obtain
\be
\varphi(X)  = \frac1{|W|}\oint \prod_{i=1}^{\rank G} \frac{dz_i}{2\pi i} \,
\frac{\prod_{\alpha: \text{ root}}\alpha(z) }{\varphi(\fg)} \, \frac{\varphi(V)}{e(V)} \, \frac{e(E)}{\varphi(E)} \;.
\ee
Then, using $\varphi(L)=x/f(x)$ for $x=c_1(L)$, we have
\be
\varphi(X) = \frac1{|W|} {f'(0)^{\rank G}}\oint \prod_{i=1}^{\rank G} \frac{dz_i}{2\pi i} \,
\frac{ \prod_{\alpha:\text{ root}} f\big(\alpha(z)\big) \, \prod_{w \in E} f\big(w(z)\big)}
{\prod_{v \in V} f\big(v(z)\big)} \;,
\ee
where $v$ and $w$ run over the weights of $V$ and $E$, respectively.
This agrees (up to a sign) with the gauge theory formula for the elliptic genus we obtained in the main part of the paper, once we use \eqref{mathelliptic}.

\section{Eta and theta functions}
\label{app: theta}

The Dedekind eta function is
\be
\eta(\tau)=q^{1/24}\prod_{n = 1}^\infty (1-q^n)
\ee
where $q = e^{2\pi i \tau}$ and $\im\tau > 0$. We will also write $\eta(q)$.  Its modular properties are
\be
\eta(\tau + 1) = e^{i \pi/12} \, \eta(\tau) \;,\qquad\qquad \eta \Big( - \frac1\tau \Big) = \sqrt{-i \tau} \, \eta(\tau) \;.
\ee
The Jacobi theta functions are
\bea
\theta_1(\tau | z)&= -i q^{1/8} y^{1/2} \prod_{k=1}^\infty (1-q^k) (1-y q^k) (1-y^{-1}q^{k-1})  \;,\\
\theta_2(\tau | z)&= q^{1/8} y^{1/2} \prod_{k=1}^\infty (1-q^k) (1+y q^k) (1+y^{-1}q^{k-1}) \;,\\
\theta_3(\tau | z)&= \prod_{k=1}^\infty (1-q^k) (1+y q^{k-1/2}) (1+y^{-1}q^{k-1/2}) \;,\\
\theta_4(\tau | z)&= \prod_{k=1}^\infty (1-q^k) (1-y q^{k-1/2}) (1-y^{-1}q^{k-1/2}) \;,
\eea
where $q$ is as before and $y = e^{2\pi i z}$. We will also use the notation $\theta_i(q,y)$.

Let us spell out some properties of the function $\theta_1(\tau|z)$ we use in the main text. Under shifts of $z$ we have
\be
\theta_1(\tau | z+a + b\tau) = (-1)^{a+b} \, e^{- 2\pi i b z - i \pi b^2 \tau} \, \theta_1(\tau|z)\label{zot}
\ee
for $a,b \in \bZ$.
Moreover
\be
\theta_1 (\tau| - z) = - \theta_1(\tau|z) \;.
\ee
The function $\theta_1(\tau|z)$ has simple zeros in $z$ at $z = \bZ + \tau \bZ$, and no poles. To compute residues it is useful to note that
\be
\theta_1'(\tau|0) = 2\pi \, \eta(q)^3
\ee
where the derivative is taken with respect to $z$. Combining with \eqref{zot}, we have
\be
\frac1{2\pi i} \oint_{u \,=\, a + b\tau} \hspace{-.5em} \dd u \; \frac1{\theta_1(\tau|u)} = \frac{(-1)^{a+b} e^{i\pi b^2\tau}}{\theta_1'(\tau|0)} \;.\label{boo}
\ee
The modular properties are:
\be
\theta_1(\tau+1| z) = e^{\pi i /4} \, \theta(\tau|z) \;,\qquad\qquad \theta_1 \Big( - \frac1\tau \Big| \frac z\tau \Big) = -i \, \sqrt{-i \tau} \, e^{\pi i z^2/\tau} \, \theta_1(\tau| z) \;.
\ee

\section{Supersymmetry and actions}
\label{app: actions}

Let us summarize here our conventions for the supersymmetry variations and the actions in Euclidean signature. For Dirac spinors we use the conventions of \cite{Benini:2012ui}: anticommuting spinors are multiplied as
\be
\epsilon \psi = \psi \epsilon \equiv \epsilon^\trans C \psi = \epsilon^\alpha C_{\alpha\beta} \psi^\beta
\ee
where $C$ is the charge conjugation matrix. We take $C = \gamma_2$ so that $C^2 = 1$ and $C^\trans = -C$, in particular $\epsilon \gamma^\mu \psi = - \psi \gamma^\mu \epsilon$.
The chirality matrix is $\gamma_3 = -i \gamma_1\gamma_2$. In components
\be
\epsilon\psi = \epsilon^+ \psi_+ + \epsilon^- \psi_- = -i \epsilon^+ \psi^- + i \epsilon^- \psi^+ \;,
\ee
hence we see how to raise and lower indices. Finally the Fierz identity for anticommuting fermions is
\be
(\bar\epsilon \lambda_1) \lambda_2 = - \tfrac12 \big[ \lambda_1 (\bar\epsilon \lambda_2) + \gamma_3 \lambda_1 (\bar\epsilon \gamma_3 \lambda_2) + \gamma_\mu \lambda_1 (\bar\epsilon \gamma^\mu \lambda_2) \big] \;.
\ee
Recall that to go to Euclidean signature we set $x^0 = i x^2$, therefore $F_{01} = i F_{12}$. Since the flux forms a holomorphic pair with the D-term in Lorentzian signature $D_L$, we define $F_{01} + i D_L = i (F_{12} + i D)$ hence $D_L = i D$.

Let us start discussing $\cN=(2,2)$ supersymmetry. First we have a vector multiplet $V_{(2,2)} = (A_\mu, \lambda, \bar\lambda, \sigma, \bar\sigma, D)$ with variations:
\bea
\delta A_\mu &= - \frac i2 \big( \bar\epsilon \gamma_\mu \lambda + \bar\lambda \gamma_\mu \epsilon \big) \\
\delta \sigma &= \bar\epsilon P_- \lambda + \bar\lambda P_- \epsilon \\
\delta\bar\sigma &= \bar\epsilon P_+ \lambda + \bar\lambda P_+ \epsilon \\
\delta\lambda &= + i \gamma_3 \epsilon \, F_{12} - \epsilon \, D - i P_- \epsilon\, [\sigma, \bar\sigma] + i \gamma^\mu P_+ \epsilon\, D_\mu \sigma + i \gamma^\mu P_- \epsilon\, D_\mu \bar\sigma \\
\delta\bar\lambda &= - i \gamma_3 \bar\epsilon\, F_{12} - \bar\epsilon\, D - i P_- \bar\epsilon\, [\sigma, \bar\sigma] + i \gamma^\mu P_- \bar\epsilon\, D_\mu\sigma + i \gamma^\mu P_+ \bar\epsilon\, D_\mu \bar\sigma \\
\delta D &= - \frac i2 \bar\epsilon \gamma^\mu D_\mu \lambda + \frac i2 D_\mu \bar\lambda \gamma^\mu \epsilon + i [\bar\epsilon P_+ \lambda, \sigma] - i [\bar\lambda P_- \epsilon, \bar\sigma] \;, \\
\eea
where
\be
P_\pm = \frac{1 \pm \gamma_3}2 \;.
\ee
With respect to the standard conventions, for instance of \cite{Witten:1993yc}, we shifted $D \to D + \frac i2[\sigma, \bar\sigma]$. Second we have a chiral multiplet $\Phi_{(2,2)} = (\phi, \bar\phi, \psi, \bar\psi, F, \bar F)$ with variations:
\bea
\delta \phi &= \bar\epsilon \psi \qquad\qquad &
\delta \psi &= i \gamma^\mu \epsilon\, D_\mu\phi + iP_+\epsilon\, \sigma\phi + i P_-\epsilon\, \bar\sigma \phi + \bar\epsilon \, F \\
\delta \bar\phi &= \bar\psi \epsilon &
\delta \bar\psi &= i \gamma^\mu \bar\epsilon\, D_\mu \bar\phi + i P_- \bar\epsilon\, \bar\phi \sigma + i P_+ \bar\epsilon\, \bar\phi \bar\sigma + \epsilon\, \bar F \\
&& \delta F &= \epsilon \big( i \gamma^\mu D_\mu \psi - i P_- \sigma \psi - i P_+ \bar\sigma \psi -i \lambda\phi \big) \\
&& \delta \bar F &= \bar\epsilon \big( i \gamma^\mu D_\mu \bar\psi - i P_+ \bar\psi \sigma - i P_- \bar\psi \bar\sigma - i \bar\phi \bar\lambda \big) \;.
\eea
The Yang-Mills Lagrangian is
\be
\cL_\text{YM} = \Tr \Big[ F_{12}^2 + D^2 + D_\mu \bar\sigma D^\mu \sigma + i D [\sigma, \bar\sigma] - i \bar\lambda \gamma^\mu D_\mu \lambda - i \bar\lambda P_+ [\sigma, \lambda] - i \bar\lambda P_- [\bar\sigma, \lambda] \Big] \;,
\ee
while the kinetic Lagrangian for the chiral multiplet is
\be
\cL_\text{mat} = D_\mu \bar\phi D^\mu \phi + \bar\phi \big( \bar\sigma \sigma + i D \big) \phi + \bar F F - i \bar\psi \gamma^\mu D_\mu \psi + i \bar\psi \big( P_- \sigma + P_+ \bar\sigma \big) \psi + i \bar\psi \lambda \phi + i \bar\phi \bar\lambda \psi \;.
\ee

Let us now move to $\cN=(0,2)$ supersymmetry. This is achieved by taking chiral parameters $P_- \epsilon = P_- \bar\epsilon = 0$. We also define complex coordinates
\be
w = x^1 + i x^2 \;,\qquad\qquad \bar w = x^1 - ix^2 \;,
\ee
so that $\gamma^w \epsilon = \gamma^w \bar\epsilon = 0$. Notice also $F_{12} = -2iF_{w\bar w}$. In the following it will be convenient to write spinors in components, in particular the SUSY parameters are $\epsilon^+, \bar\epsilon^+$.
First we have a chiral multiplet $\Phi = (\phi, \bar\phi, \psi^-, \bar\psi^-)$ with variations
\bea
\label{SUSY var chiral}
\delta \phi &= -i \bar\epsilon^+ \psi^- \qquad\qquad & \delta \psi^- &= 2i \, \epsilon^+ D_{\bar w} \phi \\
\delta \bar\phi &= -i \epsilon^+ \bar\psi^- & \delta\bar\psi^- &= 2i \, \bar\epsilon^+D_{\bar w} \bar\phi \;.
\eea
Second we have a Fermi multiplet $\Lambda = (\psi^+, \bar\psi^+, G, \bar G)$ with variations
\bea
\label{SUSY var Fermi}
\delta \psi^+ &= \bar\epsilon^+ G + i \epsilon^+ E \qquad\qquad & \delta G &= 2\, \epsilon^+ D_{\bar w} \psi^+ - \epsilon^+ \psi_E^- \\
\delta \bar\psi^+ &= \epsilon^+ \bar G + i \bar\epsilon^+ \bar E & \delta \bar G &= 2\, \bar\epsilon^+ D_{\bar w} \bar\psi^+ - \bar\epsilon^+ \bar\psi_E^- \;.
\eea
Here $\cE(\Phi_i) = (E, \bar E, \psi_E^-, \bar\psi_E^-)$ is a chiral multiplet, holomorphic function of the fundamental chiral multiplets in the theory, and it is part of the definition of $\Lambda$. Notice that $E = E(\phi_i)$ and its fermionic partner is $\psi_E^- = \sum_i \psi_i^- \, \partial E/\partial \phi_i$. Third we have a vector multiplet $V = (A_\mu, \lambda^+, \bar\lambda^+, D)$ with variations
\bea
\label{SUSY var vector}
\delta A_w &= \tfrac 12 \big( \epsilon^+ \bar\lambda^+ - \bar\epsilon^+ \lambda^+ \big) \qquad &
\delta \bar\lambda^+ &= \bar\epsilon^+ (-D-iF_{12}) \qquad &
\delta (-D-iF_{12}) &= 2\, \epsilon^+ D_{\bar w} \bar\lambda^+ \\
\delta A_{\bar w} &= 0 &
\delta \lambda^+ &= \epsilon^+ (-D+iF_{12}) &
\delta (-D+iF_{12}) &= 2\, \bar\epsilon^+ D_{\bar w} \lambda^+ \;.
\eea
Comparing with \eqref{SUSY var Fermi}, notice that the fields in the second and third column form a Fermi multiplet $\Upsilon = (\bar\lambda^+, \lambda^+, -D-iF_{12}, -D + iF_{12})$ with $\cE = 0$.

The supersymmetric action for chiral multiplets comes from the Lagrangian
\bea
\cL_\Phi &= D_\mu \bar\phi D^\mu \phi + i \bar\phi D \phi + 2\, \bar\psi^- D_w \psi^- - \bar\psi^- \lambda^+ \phi + \bar\phi \bar\lambda^+ \psi^- \\
&= -4 \bar\phi D_w D_{\bar w} \phi + \bar\phi (F_{12} + iD) \phi + 2\, \bar\psi^- D_w \psi^- - \bar\psi^- \lambda^+ \phi + \bar\phi \bar\lambda^+ \psi^- \;,
\eea
where the second equality is up to total derivatives.
For Fermi multiplets we have
\be
\cL_\Lambda = - 2\, \bar\psi^+ D_{\bar w} \psi^+ + \bar E E + \bar G G + \bar\psi^+ \psi_E^- - \bar\psi_E^- \psi^+
\ee
and for vector multiplets we have
\be
\cL_\Upsilon = \Tr \Big[ F_{12}^2 + D^2 - 2\, \bar\lambda^+ D_{\bar w} \lambda^+ \Big] \;.
\ee
Up to total derivatives, this equals the Lagrangian for the Fermi multiplet $\Upsilon$ with $\cE=0$.
Interactions are specified by holomorphic functions $J^a(\phi)$ of the chiral multiplets (and anti-holomorphic functions $\bar J^a(\bar\phi)$ of their partners), where $a$ parametrizes the Fermi multiplets in the theory:
\be
\cL_J = \sum\nolimits_a \big( G_a J^a + i \psi_a^+ \psi_J^{-a} \big) \;,\qquad\qquad \cL_{\bar J} = \sum\nolimits_a \big( \bar G_a \bar J^a + i \bar\psi_a^+ \bar\psi_J^{-a} \big) \;.
\ee
Their variation is a total derivative as long as
\be
\sum\nolimits_a E_a(\phi) J^a(\phi) = 0 \;.
\ee

All these actions are actually $\cQ$-exact. Let us define the anticommuting supercharge $\cQ$ by using commuting spinor parameters and choosing them $\epsilon^+ = \bar\epsilon^+ = 1$. The action of $\cQ$ is then immediately read off from (\ref{SUSY var chiral}), (\ref{SUSY var Fermi}) and (\ref{SUSY var vector}). We then find, up to total derivatives:
\bea
\cL_\Phi &= \cQ \big( 2i\bar\phi D_w \psi^- - i \bar\phi \lambda^+ \phi \big) \;,\qquad\qquad &
\cL_\Lambda &= \cQ \big( \bar\psi^+ G - i \bar E \psi^+ \big) \\
\cL_J &= \cQ \big( {\textstyle \sum_a} \psi_a^+ J^a \big) \;,\qquad\qquad &
\cL_\Upsilon &= - \cQ \, \Tr \big( \lambda^+ (D + iF_{12}) \big) \;.
\eea

In the reduction from $(2,2)$ to $(0,2)$ supersymmetry, the chiral multiplet $\Phi_{(2,2)}$ splits into a chiral multiplet $\Phi = (\phi, \bar\phi, P_-\psi, P_-\bar\psi)$ and a Fermi multiplet $\Lambda = (P_+\psi, P_+\bar\psi, F, \bar F)$. The vector multiplet $V_{(2,2)}$ splits into a vector multiplet $V$, with corresponding Fermi multiplet $\Upsilon = (P_+\bar\lambda, P_+\lambda, -D-F_{12}, -D+iF_{12})$, and an adjoint chiral multiplet $\Sigma = (\sigma, \bar\sigma, P_-\lambda, P_-\bar\lambda)$.
If $\Phi_{(2,2)}$ is charged under $V_{(2,2)}$, then its Fermi component $\Lambda$ has related chiral multiplet $\cE = \Sigma \Phi$ (where $\Sigma$ acts in the correct representation). It is easy to check that $\cL_\Upsilon + \cL_\Phi$ (where $\Phi$ is taken in the adjoint representation) equals $\cL_\text{YM}$, and $\cL_\Phi + \cL_\Lambda$ (where the Fermi multiplet has $\cE = \Sigma \Phi$) equals $\cL_\text{mat}$.
Superpotential interactions $W(\Phi_{(2,2)})$ become interactions $J^a(\phi) = \partial W/\partial \phi_a$.

Similarly, a $(2,2)$ twisted chiral multiplet $Y_{(2,2)}$ (which must be neutral) splits into a chiral and a Fermi multiplet. In particular the twisted chiral multiplet $\Sigma_{(2,2)}$ constructed out of $V_{(2,2)}$ splits into $\Upsilon$ (with $\cE=0$) and the chiral multiplet $\Sigma$. A twisted superpotential $\widetilde W(\Sigma_{(2,2)})$ becomes an interaction $J^\Upsilon(\sigma) = \partial \widetilde W / \partial \sigma$, and a complexified Fayet-Iliopoulos term is simply a constant $J^\Upsilon = \frac\theta{2\pi} + i\zeta$.

\bibliographystyle{ytphys}
\small\baselineskip=.93\baselineskip
\bibliography{ref}

\end{document}